\documentclass[12pt]{iopart}
\usepackage[utf8x]{inputenc}

\usepackage{amsfonts,amssymb}
\usepackage{graphicx}
\usepackage[autostyle=true]{csquotes}
\graphicspath{{./figures/}}
\usepackage[separate-uncertainty = true,multi-part-units=single]{siunitx}
\usepackage{bm} 
\usepackage[dvipsnames]{xcolor}
\usepackage{hyperref}
\hypersetup{
	colorlinks = true,
	linkcolor = NavyBlue,
	citecolor = NavyBlue,
	urlcolor = NavyBlue,}
\usepackage[nameinlink]{cleveref}	
\usepackage{orcidlink}

\Crefname{equation}{Eq.}{Eqs.}
\Crefname{figure}{Fig.}{Figs.}
\Crefname{tabular}{Tab.}{Tabs.}
\Crefname{table}{Tab.}{Tabs.}

\usepackage[numbers, compress]{natbib}
\newcommand{\cc}{\star}
\newcommand{\FT}{\mathcal{F}}
\newcommand{\conv}{\ast}

\newcommand{\boltzmann}{k_\mathrm{B}}

\newcommand{\ncol}{n_\mathrm{col}}
\newcommand{\nfit}{n_\mathrm{fit}}
\newcommand{\AW}[1]{\textcolor{black}{#1}}

\DeclareSIUnit\gauss{G}
\DeclareSIUnit\bohrradii{a_0}
\makeatletter
\newcommand{\xRightarrow}[2][]{\ext@arrow 0359\Rightarrowfill@{#1}{#2}}
\makeatother

\begin{document}
	\title[Characterizing quantum gases in time-controlled disorder realizations]{Characterizing quantum gases in time-controlled disorder realizations \AW{using cross-correlations of density distributions}}
	\author{Silvia Hiebel \orcidlink{0009-0003-8923-4216}, Benjamin Nagler \orcidlink{0000-0002-6961-0734}, Sian Barbosa \orcidlink{0009-0003-9365-7705}, Jennifer Koch \orcidlink{0009-0002-2992-3645} and Artur Widera \orcidlink{0000-0002-0338-9969}}
	\address{Department of Physics and Research Center OPTIMAS, University of Kaiserslautern-Landau, Erwin-Schrödinger-Straße 46, D-67663 Kaiserslautern, Germany}
	\ead{\mailto{widera@physik.uni-kl.de}}
	\date{\today}
\begin{abstract}
	The role of disorder on physical systems has been widely studied in the macroscopic and microscopic world. 
	While static disorder is well understood in many cases,  the impact of time-dependent disorder on quantum gases is still poorly investigated.
	In our experimental setup, we \AW{introduce and characterize a method capable of producing time-controlled optical-speckle disorder.}
	Experimentally, coherent light illuminates a combination of a static and a rotating diffuser, thereby collecting a spatially varying phase due to the diffusers’ structure and a temporally variable phase due to the relative rotation. 
	\AW{Controlling t}he rotation of the diffuser \AW{allows changing the speckle realization or, for future work,  the} characteristic time scale of the \AW{change of the speckle pattern, i.e., the correlation time}, matching typical time scales of the quantum gases investigated.
	We characterize the speckle pattern \textit{ex-situ} by measuring its intensity distribution \AW{cross-correlating different intensity patterns}.  \textit{In-situ}, \AW{we observe} its impact on a molecular Bose-Einstein condensate \AW{(BEC) and cross-correlate the density distributions of BECs probed in different speckle realizations}. 
	As one diffuser rotates relative to the other around the common optical axis, we trace the optical speckle's intensity \AW{cross-correlations} and the quantum gas' density \AW{cross-correlations}.
	Our results show comparable outcomes for both measurement methods. 
	The setup allows us to tune the disorder potential adapted to the characteristics of the quantum gas.
	These studies pave the way for investigating nonequilibrium physics in interacting quantum gases using controlled dynamical-disorder potentials.
\end{abstract}	
	\maketitle
	\section{Introduction}
	Disorder is ubiquitous and can be found in a large variety of systems ranging from, e.g., fluid dynamics \cite{Dunn.2012}, diffusion processes \cite{Sun.2020} to solids \cite{Goodrich.2014, A.M.JayannavarandN.Kumar.}, biology \cite{Jang.2015, Park.2018, BourneWorster.2019}, spectroscopy \cite{Boas.2010, Boschetti.2022}, or towards simulations for quantum computing \cite{Duncan.2017, Yin.2008, Chapman.2018}. 
	The field of ultracold gases provides a versatile tool for combining quantum gases with disordered systems \cite{Bouyer.2010}, offering a highly controllable model system for a broad range of applications. 
	Soon after the first experimental realizations of disorder potentials for ultracold gases \cite{Clement.2006}, observation of Anderson localization (AL) for quantum gases was also achieved experimentally \cite{Billy.2008}. 
	Initially predicted in the context of electron transport in crystals \cite{Anderson.1958, Lee.1985}, AL describes the absence of wave propagation in disordered media due to destructive interference of possible paths originating from multiple scattering.
	While the first observations of AL with quantum gases were made in one dimension,  localization can also occur in higher dimensions \cite{SanchezPalencia.2010, Roati.2008, White.2020, Jendrzejewski.2012, Kondov.2011, Delande.2014, Orso.2017}.
	Also, related disorder-induced observations, such as many-body localization in interacting, disordered lattice systems, have been reported \cite{Maksymov.2020, Sierant.2018, Choi.2016, Schreiber.2015, Smith.2016}. 
	Moreover, disorder has proven valuable combined with topological transport studies, such as Thouless pumping \cite{Cerjan.2020, Nakajima.2021}.\\
	
	An important recent research direction of ultracold quantum gases is the study of nonequilibrium dynamics in driven quantum systems.  
	A key asset to studying such systems is tight control over all relevant experimental parameters. 
	For dynamically-disordered systems, this implies controlling the correlation time of the disordered potential in addition to engineering the spatial correlation length.
	For disordered systems, however, most experimental realizations considered a purely static disorder, and examples of dynamical disorder studies are scarce. 
	For example,   investigations of dynamical disorder have been reported using microwaves \cite{Ruan.2020}, where the time scales of the disorder dynamics are in the millisecond range. 
	These time scales, however, are not suited for quantum gases, where relevant time scales are of the order of microseconds. 
	The impact of time-dependent disorder on quantum systems is yet largely unexplored experimentally. 
	Theoretical works predict many-body localization to resist turbulations created by time-dependent potentials \cite{Ponte.2015, Abanin.2016, Lazarides.2015} given that the driving frequency is sufficiently large. 
	Theory also predicts universal transport in dynamic disorder in the absence of localization \cite{Krivolapov.2012}. 
	At the same time,  as shown for an optical setup in \cite{Levi.2012}, AL seems to break down,  and diffusive transport or hyper-broadening is expected to occur in time-dependent disorder. 
	So far, it is unclear under which conditions localization could persist for slower dynamics.
	Furthermore, there are still open questions, e.g.,  on which timescales these processes take place and how the driving parameters influence the dynamics. \\
	Here, we show a simple scheme to create time-dependent optical disorder potentials where the spatial statistical properties of the potential are time-independent. 
	To analyze \AW{the disorder properties of this scheme,  we here experimentally measure the optical intensity cross-correlations, relating intensity distributions of two different disorder realizations,  which decay as the disorder realization is experimentally changed in discrete steps.
	We address the extent to which these optical cross-correlations are reflected by the cross-correlations of quantum gas densities measured in different static disorder realizations. }
	While the former is a direct measurement of the disordered optical potential's cross-correlation, the latter is influenced by several effects. Theoretically, the density distribution of a BEC in disorder due to interaction effects has been studied; see Ref.~\cite{SanchezPalencia2006}. Experimentally, we address the influence of column integration, imaging noise,  inhomogeneities, and finite temperature in images of  quantum gases and whether the \AW{cross-correlations between density distributions taken in different realizations }of the underlying disorder potential can be deduced.
	We find that, indeed, the atomic density \AW{cross-}correlations between two absorption images of quantum gases trapped in different optical disorder realizations resemble the direct optical correlations.
	This result will allow the study of quantum gases in dynamical disorder when the optical speckle correlation is continuously reduced during the trapping of a quantum gas in future work.

	\section{Static and dynamic optical speckle}
		\begin{figure}[t]
			\centering
			\includegraphics[width=.6\textwidth, trim=110 60 140 80, clip]{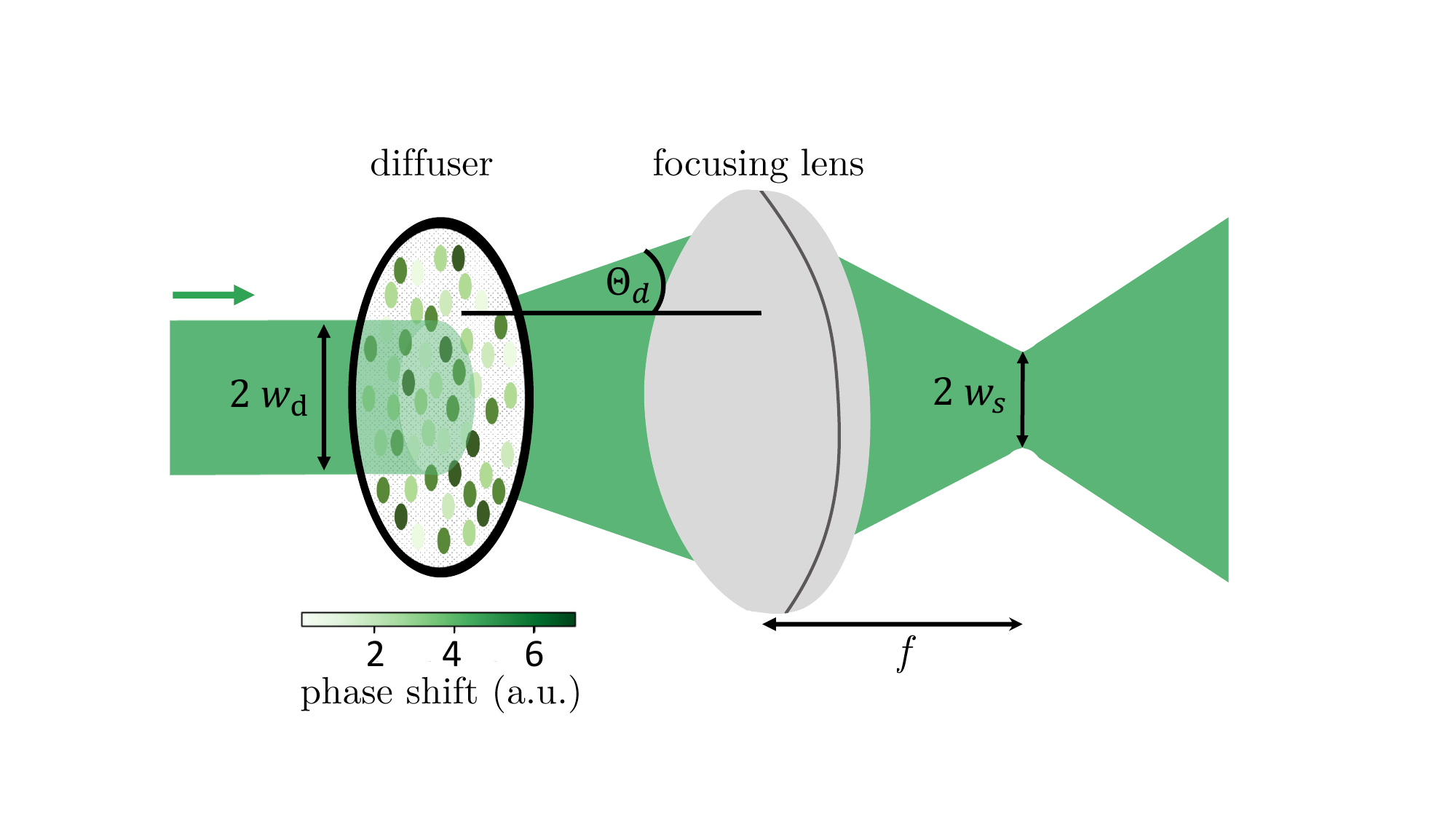}
			\caption{Schematic drawing of a simple diffuser setup consisting of a diffuser and a focusing lens. A collimated Gaussian laser beam with waist $w_\mathrm{d}$ illuminates the diffuser, which imprints random amplitudes and phases onto the light. The beam is diffracted into a cone with a half-opening angle $\Theta_\text{d}$. Subsequently, an objective lens focuses on it with a focal length of f. The beam has a Gaussian-shaped intensity distribution envelope with the waist $w_s$ in the focal plane.}
			\label{fig:diffuser_lens}
		\end{figure}
	
	The prime example of random disorder phenomena is optical speckles. 
	Speckles occur as interference fringes of random scattering at rough surfaces or within diffusive media \cite{Dainty.1975}. 
	Applications include medical imaging \cite{Zakharov.2009}, optical coherence tomography \cite{Schmitt.1999} or astronomy \cite{Goodman2007}, but also forming disordered potentials for ultracold gases \cite{Shapiro.2012, Falco.2010, Yue.2020}. 
	For quantum gas applications, optical speckle patterns can conveniently be realized by transmitting a laser beam through a diffuser, e.g., a ground glass plate. This simple approach has evolved to an established method for optical disorder production \cite{Dainty.1975, Francon.1979, Wang.2018}.  Microscopically, the surface structure acts as a random collection of scattering centers imprinting a random amplitude and phase distribution onto the light field as shown in \Cref{fig:diffuser_lens}. 
	
	When a collimated Gaussian-shaped laser beam with waist $w_\mathrm{d}$ is transmitted through a diffusive plate, the light gets diffracted at every point of the intensity profile, and an angular distribution is formed with cone angle  $\Theta_\mathrm{d}$.
	An imaging lens with focal length $f$ creates a disordered intensity distribution in the focal plane. This distribution in the imaging plane consists of individual speckle grains and has a waist of 
	\begin{equation} \label{eq:diffusive_angle}
		w_\text{s}=\Theta_\mathrm{d} f .
	\end{equation}
	For the application as an optical potential, especially the local spatial and temporal properties of the speckle are \AW{important} \cite{Falco.2010}.
	The transversal correlation length $\sigma_\mathrm{s}$ of the speckle measures the average size of the speckle grains in the image plane or, equivalently, the length scale on which the speckle preserves its coherence.
	In the focal plane of a lens, the grain size is minimized in the transversal direction.  Here, the imaging resolution limits the achievable correlation length, which can be a few hundred nanometers.
	The speckle-grain extension in the longitudinal direction, i.e., along the propagation axis, is much larger and determined by the Rayleigh length of the focusing system.
	Experimentally, the speckle's correlation length can be determined from arbitrary light distributions by calculating each dimension's angle autocorrelation function of the speckle pattern.  
		
\begin{figure*}
			\centering
			\includegraphics[trim=0 19 28 10,clip]{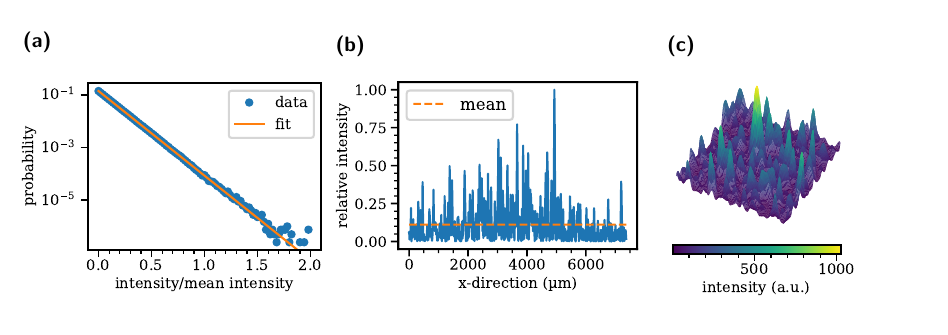}
			\caption{Defining speckle characteristics. (a) intensity-probability distribution for a recorded speckle image as an absolute count rate. The orange line depicts an exponential fit to the semi-logarithmic plotted measurement data. (b) One-dimensional intensity distribution for a speckle image with the mean value indicated as an orange dashed line. (c) 3d image of a typical speckle pattern of size $(384 \times 384)\, \SI{}{\micro \meter}$, where the speckle's correlation length is $(575 \pm 15)\,\SI{}{\nano \meter}$.}
			\label{fig:speckle_properties}
		\end{figure*}

	Knowing the magnitude of the correlation length is essential for using the speckle as a dipole potential for quantum gases.  For a Bose-Einstein condensate, a relevant length scale is the healing length \cite{pethick_smith_2008} $\xi=1/\sqrt{8\pi n_0 a}$ of the gas in the trap center with peak density $n_0$ of the gas and s-wave scattering length $a$. The healing length represents the smallest length scale on which the condensate's wave function can react to perturbations. 
	In our case,  the correlation length $\sigma_\mathrm{s}$ is larger than the healing length	$\xi$  by a factor of two, allowing the condensate to resolve all details of the speckle. 
	Since the speckle imposes a potential onto the atomic cloud, it imprints a density modulation on the condensate. 
	We can infer information about the speckle potential by recording the column densities of the condensate \cite{SanchezPalencia2006, Falco.2007}.
		Furthermore, for an ideal, fully-developed speckle, the distribution of intensities is the same for all realizations \cite{Dainty.1975}.  It is given by the negative exponential intensity-probability-distribution \cite{Goodman2007}
		\begin{equation}
			P(I)=\frac{1}{\langle I \rangle} \exp \left( - \frac{I}{\langle I \rangle}\right)
		\end{equation}
		with intensity $I=I(\vec{r})$ at the point $\vec{r}$.
		This leads to the fact that high-intensity speckle grains are scarce. The average intensity of a speckle realization is relatively low. 
		A  typical measured intensity-probability distribution and an intensity distribution for one column of a typical speckle image are shown in \Cref{fig:speckle_properties}.
		
\begin{figure}[b]
		\centering
		\includegraphics[width=0.6\textwidth]{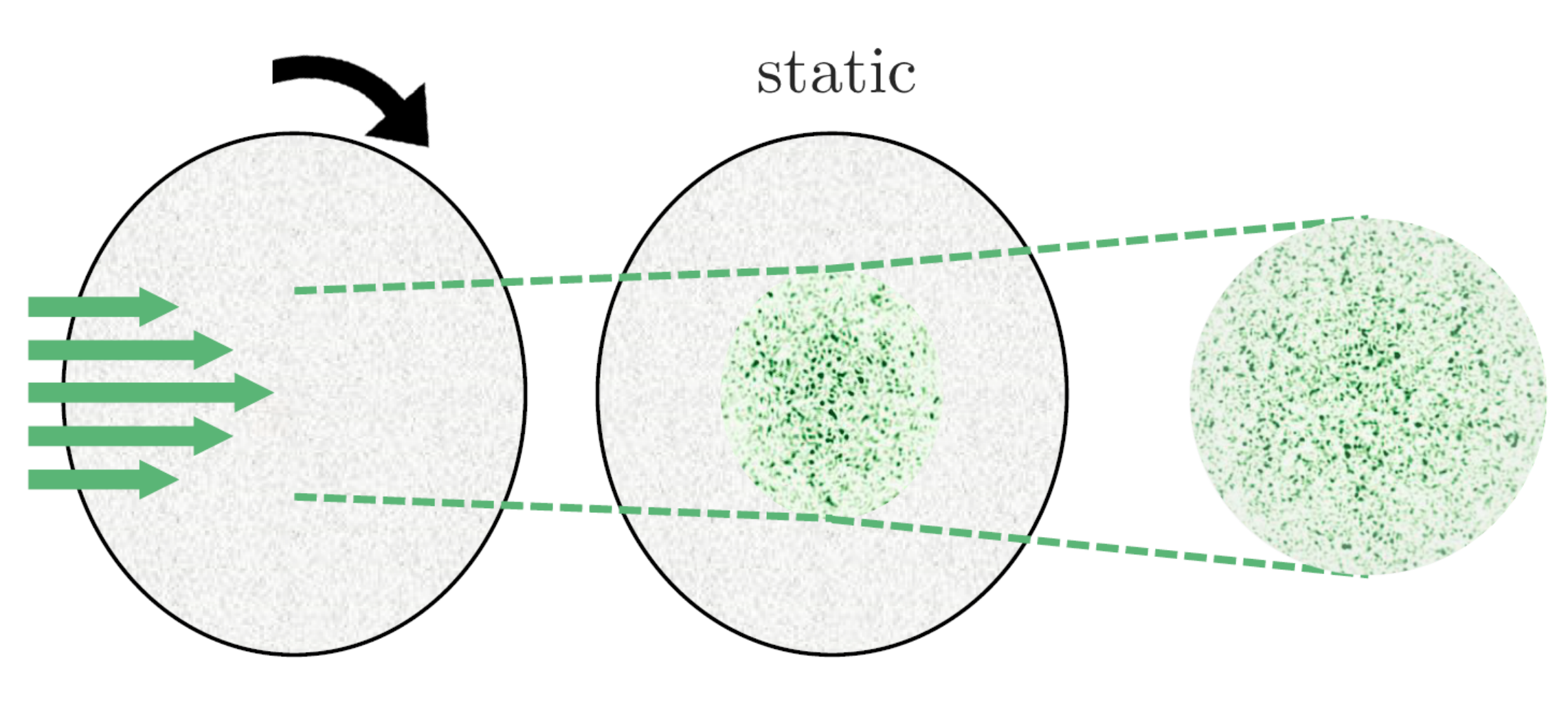}
		\caption{Sketch of the diffuser setup. Coherent light with a wavelength of $\lambda=\SI{532}{\nano\meter}$ is transmitted through a rotating diffuser, which widens the beam and imprints a random and rotating intensity distribution onto the beam. That illuminates a second diffuser, again imprinting a random phase and amplitude distribution. Combining two diffusers in a row leads to a dynamically changing disorder pattern.}
		\label{fig:diffuser_setup}
		\end{figure}

		The properties considered so far hold for speckle patterns in general. In our experiment, the speckle pattern can be applied as a static potential, or \AW{be changed in time}. 
		To \AW{this end,} speckle patterns can, among others, be manipulated by modifications of the diffusive medium. 
		In our case, the speckle is created with the help of an optical diffuser. 
		The speckle pattern can change, e.g., by movements or rotation of the diffusive medium \cite{Shi:19}, which causes decorrelation effects. 
		However, the entire speckle pattern is shifted or rotated for a simple translation or rotation.\\
		In our setup, we supplement the scheme shown in \Cref{fig:diffuser_lens} by an identically-constructed second diffuser (Edmund Optics \#47-991) in a distance of up to \SI{10}{\centi \meter} to create a speckle pattern that can be \AW{changed via the relative angle between the diffusers} (\Cref{fig:diffuser_setup}). Each diffuser plate imprints a random phase distribution onto the incoming laser beam. The first diffuser rotated around the principal  optical axis, produces a rotating speckle pattern onto the second diffuser. This, in turn, imprints another random phase- and amplitude distribution onto the rotating distribution. This combination leads to a disordered interference pattern that can change depending on the relative rotation angle of the two diffusers around the principal optical axis. That changes the position of the speckle grains and their intensity. As a side effect, the second diffuser also widens the beam further, slightly reducing the transmitted light's intensity.
\AW{		The speckle pattern can change stepwise by rotating one of the diffusers for a specific angle or dynamically by rotating the diffuser with a certain angular velocity.
		In the present work, we use a stepwise change between different realizations of one experimental run for characterization purposes. The continuous, dynamic change enabled by this experimental scheme allows us to study quantum gases in dynamical disorder, which is beyond the scope of this work.}
		
		To characterize the speckle pattern produced independently from the quantum gas, a test setup was built that simulated the characteristics of our quantum gas setup. For details on the main setup, see Refs.\cite{Gaenger2018, Nagler2020}. 
		The test setup includes a laser beam of wavelength $\lambda = \SI{532}{\nano\meter}$  illuminating the abovementioned diffuser setup  with waist $w_d$, see \Cref{fig:diffuser_lens}. The speckle pattern results from imaging this randomized phase distribution using a focusing lens with a numerical aperture of 0.3. The image plane corresponds to the position of the molecules in the quantum gas setup.  A microscope (Olympus RMS20X and Thorlabs TTL180A) with a magnification factor of 20 maps the pattern on a CCD camera (FLIR Blackfly BFS U3-63S4M-C), which is used to record and characterize the optical intensity distribution. \\
		A diffuser itself brings its own correlation length $\sigma_\mathrm{d}$, which is given by its surface structure. 
		It can be calculated as \cite{phdrichard}
		\begin{equation}
			\Theta_d=\frac{\lambda}{\pi \sigma_\mathrm{d}} \label{eq:corr_trans}
		\end{equation}
		with wavelength $\lambda$ of the incoming light. 
		Including a second diffuser to the setup leads to a modified correlation length $\sigma_\mathrm{D}$ of the whole setup compared to just using one diffuser in \Cref{eq:corr_trans}.  
		The correlation length for configurations with two diffusers of different divergence angles can be assumed in a simple model as
		\begin{equation}
			\sigma_\mathrm{D}=\sqrt{\sigma_\mathrm{d,1}^2+\sigma_\mathrm{d,2}^2}=\frac{\lambda}{\pi}\sqrt{\frac{1}{\Theta_\mathrm{d,1}^2}+\frac{1}{\Theta_\mathrm{d,2}^2}} \label{eq:divergence_angles}
		\end{equation}
		with the compound correlation length $\sigma_\mathrm{D}$, the single correlation lengths $\sigma_\mathrm{d,i}$ corresponding to each plate, wavelength $\lambda$ and the half divergence angles $\Theta_\mathrm{d,i}$.

	\section{Ex-situ characterization of the correlation angle} \label{sec:exsitu}
		\begin{figure*}
		\includegraphics[width=0.9\textwidth]{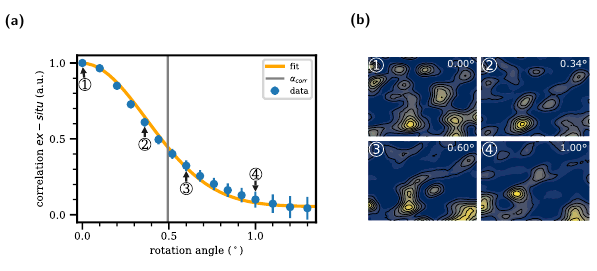}
		\caption{Changes in the speckle patterns for diffuser rotation. (a) The maximum value of the cross correlation function of the speckle intensity distribution for different rotation angles with two diffusers of a \ang{1} divergence angle each. The blue points show the measurement data. Error bars mark the uncertainty of a fit to extract the maximum value from the \textit{ex-situ} correlation. A solid orange line shows a Gaussian fit curve to the data points. The gray line marks the correlation angle at which the autocorrelation has dropped to the half maximum. bars denote the fitting error. (b) Contour plot of an extract of the resulting speckle pattern for the measurement points depicted in (a) with rotation angles of \ang{0.00}, \ang{0.34}, \ang{0.60}, and \ang{1.00}.}
		\label{fig:cctwoplates}
	\end{figure*}
	
	As a first speckle characterization, we measure the speckle's transversal correlation lengths of the intensity distribution produced by the combined diffuser setup and characterize the grain size. 
	We computationally analyze the autocorrelation of the speckle patterns, yielding a Gaussian decay of correlation in each direction. The transversal correlation length can be extracted by fitting a Gaussian function through the center of the two-dimensional angle autocorrelation function. The full width at half maximum of that function measures the speckle's grain size.
	These measured correlation lengths are not necessarily equal in the two directions of an image because of the inhomogeneities of the camera sensor, the imaging system, and the speckle itself. In our case, the correlation length was determined to be $\sigma_{\mathrm{s}_x} =$ \SI{558.7 \pm 3.6}{\nano\meter} in the $x$ direction, and $\sigma_{\mathrm{s}_y} =$ \SI{554.9 \pm 3.6}{\nano\meter} in the $y$ direction. We used the method described in \cite{Gaenger2018} for the determination.

	With our setup, we can, in particular, provide and control the decorrelation of optical speckles. Experimentally, this decorrelation is realized by the rotation of one of the diffusers \AW{by a given rotation angle}, and it manifests itself in a change of the speckle pattern. 
	However, specific details of the experimental setup will influence the particular decorrelation behavior.
	We determine different influencing factors and their impact on the speckle pattern by varying parameters in the experimental setup, including the beam waist illuminating the diffuser and the distance between the diffusers. 
	For each of the following measurements in this section, the first diffuser is rotated in steps between \ang{0.04} and \ang{0.10} while the other is kept static. 
	This method allows evaluating the most suitable parameters for later use in the quantum gas setup.\\
	The decorrelation can be quantified by numerically cross correlating pictures taken of the initial optical intensity pattern and after rotation of the diffuser by a given angle. 
	This function is given by
	\begin{equation}\label{eq:correlation-function}
		C_\Theta ( \vec{\delta r} ) = \langle I_{\Theta=\ang{0}} (\vec{r}) I_\Theta(\vec{r}+\vec{\delta r}) \rangle
	\end{equation}
	with rotation angle $\Theta$, the speckle intensity distribution $I_\Theta$ and position vector $\vec{r}=(x,y)$.  \Cref{eq:correlation-function} therefore quantifies the change between two speckle patterns differing by the rotation angle $\Theta$.
	In the image plane, the speckle pattern changes with rotation, and new peaks in the optical intensity appear while others move and eventually vanish. 
	This is shown in \Cref{fig:cctwoplates} (b) for rotation angles of $\ang{0.00}, \ang{0.34}, \ang{0.60}$ and $\ang{1.00}$. Plotting the maximum of the \AW{intensity} cross-correlation versus the rotational angle results in a typical decorrelation curve \cite{Shi:19} having the shape of a Gaussian function as exemplified in \Cref{fig:cctwoplates}. 
	This yields a characteristic angle at which the correlation has decayed to half its maximum (grey vertical line). That angle we define as the (rotational) correlation angle. \\
	Details of the illuminating beam and the diffusers determine this angle. To quantify this connection with a simple model, we use a geometric approach: In our setup, a rotation of the first diffuser by a small angle $\alpha$ causes a local displacement of
	\begin{equation}
		\Delta s = w_\mathrm{d}\cdot \alpha .
	\end{equation}
	If this shift is larger than the correlation length of the other diffuser, the amplitude and phase distributions and, therefore, the whole speckle pattern behind the two diffusers is significantly changed, hence decorrelated. \\
	For a phase distribution shift $\Delta s > \sigma_\mathrm{d} $, this leads to a correlation angle 
	\begin{equation}\label{eq:sigma_waist}
		\alpha_{corr}\propto \frac{\sigma_\mathrm{d}}{w_\mathrm{d}}.
	\end{equation}
		In the remainder of this work, the correlation angle will be changed in discrete steps, and molecular BECs will experience a static disorder potential.  

\begin{figure*}
		\includegraphics[width=1\textwidth]{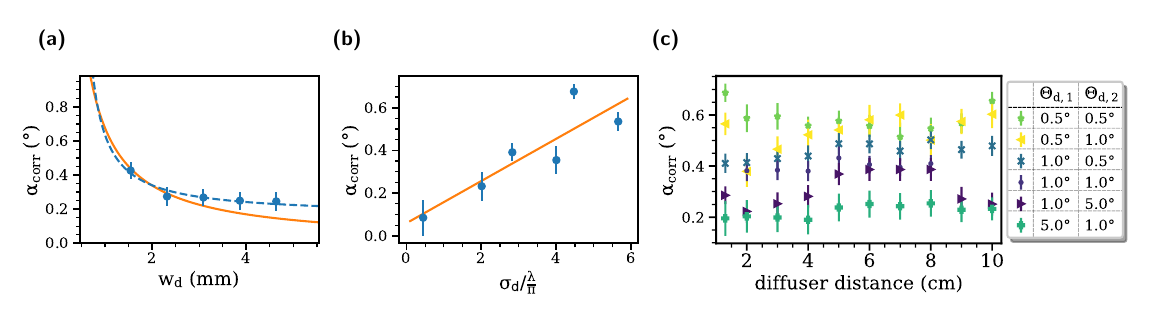}
		\caption{Influence of different setup changes on the correlation angle. Errors bars correspond to fitting errors. (a) Changes in the diameter of the incoming laser beam $w_\mathrm{d}$ on the diffusers. The orange line depicts a theoretical course $\frac{\theta_\text{D}}{2 \cdot w_\text{d}}$, the blue dotted line a 1/x-fit. (b) Influence of the correlation length $\sigma_\text{d}$. The orange line depicts a linear fit to the data points. (c) Influence of the distance between the two diffusers. Measurements for different combinations of diffusers are drawn.}
		\label{fig:influencingfactors}
	\end{figure*}

	The decorrelation can be precisely tuned by experimentally modifying the beam diameter of the laser beam passing through the diffusers as shown in \Cref{fig:influencingfactors}. 
	For example, the distance between the two diffusers and the diffusive angle will influence the correlation angle according to \Cref{eq:diffusive_angle} and \Cref{eq:corr_trans}.
	Intuitively, the scattering of the first diffuser will lead to an additional divergence of the laser beam, modifying the beam diameter on the second diffuser depending on the distance. 
	While this widening is hardly significant for small diffuser distances, it must be considered for large distances.
	The correlation angle as a function of beam diameter is shown in \Cref{fig:influencingfactors} (a).  
	Additionally, the expectation from equation \Cref{eq:sigma_waist} is shown as solid orange line with $\frac{\pi \cdot \sigma_\text{d}}{w_\text{d}}$. 
	The curves are in good agreement.
	However, the diffusers' divergence angles are another free parameter, see \Cref{eq:divergence_angles}.
	The dependence  of the correlation angle on $\sigma_\mathrm{d}$, i.e., the combined correlation length of the two diffusers, is presented in  \Cref{fig:influencingfactors}(b) for different combinations of diffusers with divergence angles of \ang{0.5}, \ang{1.0}, \ang{5.0} and \ang{10.0} (Edmund Optics \#47-991). 
	Here, the diffuser with the smaller divergence angle is placed first in the beam path for each combination.
	The resulting change in the correlation angle with combined correlation length is linear, indicated by the solid orange line in \Cref{fig:influencingfactors}(b). 
	Finally, the influence of the distance between the diffusers on the correlation angle is studied. 
	As shown in \Cref{fig:influencingfactors} (c), the combination of different diffusers mainly influences the correlation length.
	The correlation angle stays the same within the errors for the distance in question. The combinations with smaller divergence angles, especially the combination of two \ang{0.5} divergence angles, lead to larger correlation angles. Moreover, the order of the two diffusers plays a role if diffusers with different divergence angles are used. 
	If the plate with a smaller divergence angle is placed in front, this leads to a larger correlation angle.
	
	In the following part for measurements with a molecular BEC, the distance between the two diffusers is approximately \SI{5.5}{\centi \meter}. This is sufficient for optical access to the system. Two diffusive plates with divergence angles of $\ang{1.0}$ (Edmund Optics \#47-991) are chosen to compromise a large correlation angle and the desired beam width after the diffuser setup. With the beam waist given in the quantum gas setup approximately \SI{4}{\milli\meter}, the test setup suggests correlation angles of roughly \ang{0.5} -- \ang{0.6}.\\

	\section{Preparation of a molecular Bose-Einstein condensate}
		We produce a molecular BEC comprising fermionic $^6$Li atoms prepared in the two lowest-lying hyperfine substates of the electronic ground-state $^2S_{1/2}$ \cite{Gaenger2018}. 
		The experimental sequence starts with laser cooling of atoms in a Zeeman slower and subsequently in a three-dimensional magneto-optical trap.		
		After laser cooling, we evaporatively cool the sample at an external magnetic field of \SI{763.6}{\gauss}, close to a broad Feshbach-resonance centered at \SI{832.2}{\gauss}, which we use to  tune the inter-particle scattering length.
		At the end of the evaporation, the final dipole trap power of  \SI{10}{\milli \watt}, corresponding to a trap depth of $k_B \times $ \SI{500}{\nano \kelvin},  is held constant for \SI{100}{\milli \second}, producing approximately N $\approx \num{4.3e5}$  bosonic molecules at temperatures of about \SI{50}{\nano \kelvin}.
		Here, the \textit{s}-wave scattering length lies at ${a=\SI{2706}{\bohrradii}}$ \cite{Zurn.2013}, resulting in a chemical potential of ${\mu=\SI{250}{\nano\kelvin}\times\boltzmann}$ and a healing length of ${\xi=\SI{270}{\nano\meter}}$, which is smaller than the shortest correlation length of the speckle.
		Hence, the wave function of the molecular BEC resolves the microscopic details of the potential~\cite{SanchezPalencia2006}.
		The trapping frequencies of the combined optical and magnetic traps are $(\omega_x,\omega_y,\omega_z)=2\pi\times{(\SI{195}{\hertz},\SI{22.6}{\hertz},\SI{129}{\hertz})}$ with axial confinement in $y$ direction by a magnetic saddle potential.
		The many-body relaxation time is $h/\mu \approx $ \SI{190}{\micro \second}. 
		
		After preparing the molecular cloud, we apply the optical speckle pattern discussed above, serving as a random potential for the molecules. 
		The speckle pattern here is created with a Gaussian laser beam at a wavelength of \SI{532}{\nano \meter} as in the test setup, using a power of up to \SI{10}{\watt}. We use the same ground-glass plates as tested before and insert one in a high-precision rotation stage (Owis DRTM65-D35-HiDS), providing rotation velocities up to $\ang{2100}/s$, as well as discrete steps with a minimal rotation angle of $\ang{0.036}$. The other diffusive plate is kept static. 
		This combination provides a controllable dynamic disorder potential, which is focused on the position of the molecules. 
		The \AW{static} speckle potential is slowly introduced during \SI{50}{\milli\second} of linear ramping to minimize the creation of excitations in the system. 
		Subsequently, it is held for \SI{100}{\milli \second}. The gas is trapped in the combined magnetic trap and speckle potential to allow the molecules to react to the disorder potential and equilibrate. 
		Finally, we image the cloud \textit{in-situ} to infer the column density distribution via resonant high-intensity absorption imaging.  
		\AW{Before the next quantum gas is prepared, the diffuser is rotated by a given angle.}
		Importantly, during every experimental realization, the molecular BEC is trapped in a static disorder realization, and a discrete angle changes the specific realization before the next quantum gas is prepared. 
		Studying the evolution of quantum gases in continuously changing disorder, i.e., when the disorder realization and hence the disorder correlation changes while the molecular BEC is trapped, is beyond the scope of this work.
		
	\section{In-situ measurement of speckle waist}
\begin{figure*}
			\centering
			\includegraphics[trim=0 19 28 10,clip]{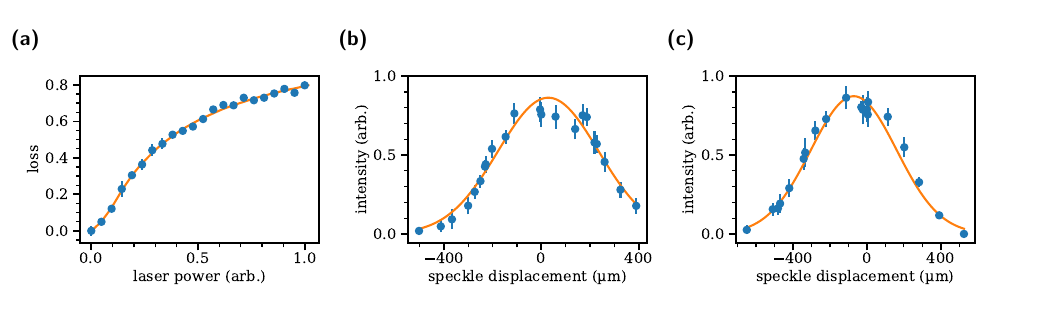}
			\caption{Determination of speckle waist and alignment. (a) Losses in the molecular cloud versus speckle laser power. Error bars denote the standard deviation of five repetitions. The laser power measures the speckle intensity with the speckle potential centered at the molecular cloud. The solid line is a fifth-order polynomial function fitted to the data. (b) and (c) Disorder intensity envelope along two orthogonal directions in the focal plane of the objective. The speckle pattern is shifted with the same laser power, and the intensity distribution is acquired via the mapping from (a). The error bars show the standard deviations for five repetitions. The resulting intensity distribution is fitted with a solid orange line as a Gaussian function, which gives waists of (b) $466 \pm 25$ \SI{}{\micro \meter} and (c) $414 \pm 25$ \SI{}{\micro \meter}.}
			\label{fig:speckle_waist}
		\end{figure*}

		Inside the vacuum chamber, the magnetic field center determines the exact cloud's position. 
		Ideally, this position should coincide with the center of the speckles' Gaussian envelope. 
		To ensure both speckle and trap are adequately overlapped, we \AW{employ the loss of molecules upon quickly switching on the speckle, which depends on the local strength of the disorder potential. We }characterize the loss of molecules depending on the speckle's exact positioning and the laser power creating the speckle.
		For the adjustment, a tiltable mirror between the diffusers and the objective is used to slightly alter the angle along both available directions at which the speckle beam enters the objective. 
		With this procedure, we aim at the largest-possible molecule losses the speckle can induce when the speckle potential is suddenly switched on. 
		Different from the usual measurement sequence, we switch the speckle rapidly on a time scale of less than \SI{1}{\micro\second}, and the free-evolution time in the speckle is \SI{400}{\milli\second}. 
		For maximum losses, the disorder potential is centered on the molecular cloud. 
			Once this optimal adjustment has been found, we map the fraction of molecules lost to the total power of the speckle pattern as shown in  \Cref{fig:speckle_waist}~(a). 
		With this mapping, we can measure the waist of the disorder potential by shifting the relative position of speckle and cloud positions along two orthogonal directions via the objective; the position-dependent losses for constant speckle power as shown in \Cref{fig:speckle_waist}~(b) and (c) reflect the Gaussian envelope of the speckles' intensity distribution. 
		 We extract the beam waist by fitting the data taken with a Gaussian curve. 
		We find waists of $466 \pm 25$ \SI{}{\micro \meter} in one direction and $414 \pm 25$ \SI{}{\micro \meter} in the  orthogonal direction.
		 The speckle size is thus large enough for the molecular cloud with an extension of $\sim$ \SI{200}{\micro \meter} on its long axis.

	\section{In-situ measurement of the correlation angle} 		\label{subsec:density_corr}
		In order to study the influence of the optical speckle potential on a quantum gas, we apply the speckle pattern and adjust it to the position of the molecular BEC as described above; for details, see \cite{Nagler20202}.
		Focused onto the molecular cloud in the center of a vacuum system, the speckle pattern can not be characterized directly but only via its impact onto the ultracold cloud via the dipole force. 
		However, detailed \textit{in-situ} knowledge about the local speckle properties is crucial beyond the global properties, such as its waist, because small changes in the optical imaging system might lead to a significant change of, e.g., the correlation length.\\
		We explain the reasons for our assumption that optical and density \AW{cross-correlation and hence correlation angle} are comparable in detail in the Appendix \ref{sec:dens_corr}.
		\\		
		\begin{figure}
			\centering
			\includegraphics[trim=0 3 20 5,clip]{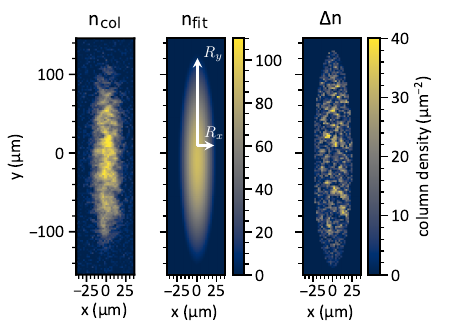}
			\caption{Absorption image processing.  Each recorded absorption image $\ncol(x,y)$ (left) is fitted with a smooth, two-dimensional Thomas-Fermi profile $\nfit(x,y)$ (center), which drops to zero at the indicated Thomas-Fermi radii $R_{x/y}$.  Subtraction of the fitted profile from the measured column density distribution yields the speckle-induced density fluctuations $\Delta n(x,y)=\ncol(x,y)-\nfit(x,y)$ (right), pixel values outside the Thomas-Fermi profile are set to zero in $\Delta n$. }
			\label{fig:cloud_evaluation}
		\end{figure}				
		
\begin{figure*}
			\centering
			\includegraphics[trim=0 3 0 0,clip, width=1.0\textwidth]{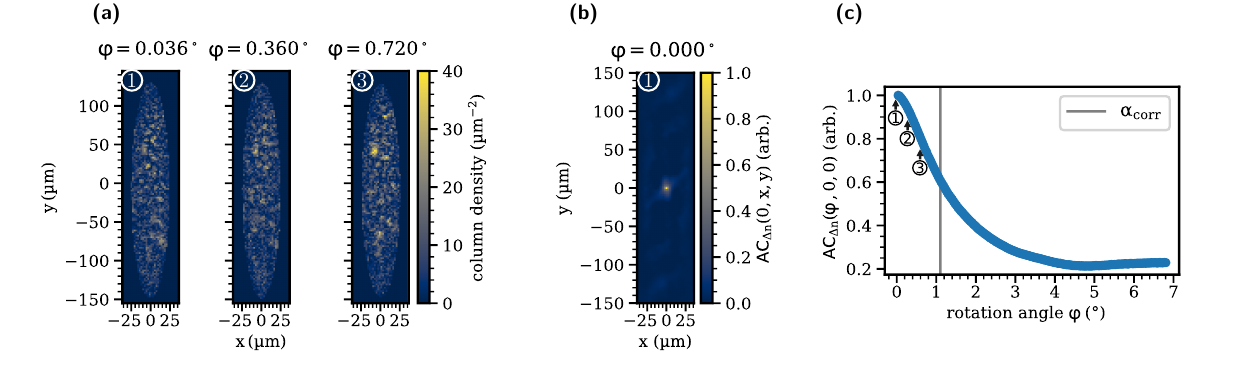}
			\caption{\textit{In-situ} measurement of the correlation angle. (a)~Speckle-induced density fluctuations for different rotational angles of the diffuser.  Small rotations ($\ll \ang{1}$) already lead to significant changes in the difference images.  (b)~Angle autocorrelation $\mathrm{AC}_{\Delta n}(\delta \varphi=0,x,y)$ calculated from the column density distribution $\ncol(\varphi,x,y)$ for the initial rotation angle with $ \varphi = \ang{0}$. Without rotation, corresponding to the left image from (a), it shows one significant peak in the middle of the image. With increasing rotation angle, the angle autocorrelation function's peak height decreases.
			(c)~Excerpt of angle autocorrelation function calculated from the first 192 column density distributions
			with ${\varphi=\ang{0},\dots,\ang{6.876}}$. As the angle of rotation increases, the speckle pattern changes, and with it, the height of the autocorrelation maximum decreases in a Gaussian manner and approaches a finite value.  This offset is due to random residual correlations between the images. The numbers indicate the corresponding values to the images from (a). The gray line marks the correlation angle of \ang{1.14}, at which the correlation function has dropped to half its initial value.}
			\label{fig:correlation_angle_bec}
		\end{figure*}

		To compare the correlation properties inferred from molecular density distributions to the values obtained from optical measurements, we prepare molecular BECs described above and expose them to the disorder potential, ramped up linearly within \SI{50}{\milli\second}.  
		After reaching the final disorder strength of ${\SI{646.5}{\nano\kelvin}\times\boltzmann}$, the cloud is held in the speckle for \SI{100}{\milli\second}, allowing the system to equilibrate. 
		Finally, the sample's column density distribution $\ncol(x,y)$ is measured using high-intensity-corrected absorption imaging~\cite{Reinaudi2007}. 
		This procedure is repeated 380 times, where in-between repetitions, the angle $\varphi$ of the second diffuser is increased in steps of \ang{0.036} up to a rotational angle of $\ang{13.640}$. 
		For identical parameters, we repeat these measurements four times.
		That yields the column density distribution $\ncol(\varphi,x,y)$ for each angle $\varphi$. 
		One exemplary absorption image is shown in \Cref{fig:cloud_evaluation}. 
		We first fit the absorption images of the disturbed clouds with a two-dimensional Thomas-Fermi profile $\nfit(x,y)$. Fitting the corresponding molecular cloud ensures to avoid negative influences of shot-to-shot fluctuations compared to subtracting the density profile of a cloud without the speckle potential. The difference between this fit and the measured distribution yields the speckle-induced density fluctuations $\Delta n(x,y)=\ncol(x,y)-\nfit(x,y)$ quantifying the speckle patterns' influence. 
		The influence of the diffusers' relative angle is shown in \Cref{fig:correlation_angle_bec}~(a), showing the position-resolved fluctuations $\Delta n (x,y)$ for the three rotation angles of $\varphi=0.00^\circ, \varphi=0.36^\circ,$ and $\varphi=0.72^\circ$. 
		Density maxima from the first picture move or vanish as the diffuser rotates to larger angles. In contrast, new maxima appear in other positions, as seen in the \textit{ex-situ} measurements. 
		Although the basic structure of the original pattern can still be seen in the third image ($\varphi=\ang{0.72}$), several differences have developed. 
		With further rotation of the diffuser, these differences increase. 
		Only random correlations to the initial pattern are found for rotation angles larger than $\ang{4.5}$.
		We quantify these differences between images for each rotation angle $\varphi$ to the initial one via the autocorrelation \AW{within one image (for $\delta \varphi =\ang{0}$) or the cross-correlation between two images (for $\delta \varphi \ne\ang{0}$)} in two-dimensional space $x$ and $y$ and angle $\varphi$.  \Cref{fig:correlation_angle_bec}~(b) shows the molecular cloud's position-resolved autocorrelation of the initial cloud (i.e.,  $\varphi=0^\circ$). 
		The central ($x=0=y$) auto-correlation peak decreases height with increasing rotation angle. 
		We normalize the peak height of the different images to this initial peak height of the autocorrelation function.
		The normalized autocorrelation function in the angle and space domain is given by
		\begin{equation}
			\mathrm{AC}_{\Delta n}(\delta \varphi,\delta x,\delta y) =  \frac{\sum_{(x_i,y_j)} \Delta n(\varphi,x_i,y_j)\Delta n(\varphi+\delta \varphi,x_i+\delta x,y_j+\delta y)}{\sum_{(x_i,y_j)} \Delta n(\varphi,x_i,y_j)^2},
		\end{equation}
		where the summation runs over all pixel $(x_i,y_j)$ of the recorded image.
		\AW{We note that for $\delta\varphi=0$, the autocorrelation of a single image is formed (\Cref{fig:correlation_angle_bec}~(b)). At the same time, for $\delta \varphi \neq 0$, the cross-correlation between two density distributions trapped in different speckle realizations is computed (\Cref{fig:correlation_angle_bec}~(c)).}
		The power spectral density (PSD) in \textit{k}-space can be calculated via the Fourier transformed density of the difference images with \cite{MILLER2004369}
		\begin{equation}
			\mathrm{PSD}(k_x,k_y) = \left|\mathcal{F}(\Delta n(\varphi,x,y))\right|^2
		\end{equation}
		With this, we can calculate the autocorrelation function, which is proportional to the inverse Fourier transformation of the PSD
		\begin{equation}
			\mathrm{AC}_{\Delta n}(\delta \varphi,\delta x,\delta y) \propto \mathcal{F}^{-1}(\mathrm{PSD}(k_x,k_y))
		\end{equation}
		From the decay of the maximum of each of these autocorrelation functions $\mathrm{AC}_{\Delta n}(\delta \varphi,0,0)$ as a function of the relative rotation angle $\delta \varphi$ of the diffusers, the correlation angle can be determined, see \Cref{fig:correlation_angle_bec}~(c).
		
		We again define the correlation angle as the angle where the \AW{cross-}correlation has decayed to half the initial value at an angle of $ \varphi \approx \ang{1.14}$.
		This angle differs from the correlation angle determined \textit{ex-situ} but is in the same order of magnitude. A combination of experimental limitations may explain the difference.
		The first constraint originates from limits in absorption imaging.
		We can only take two-dimensional images of the actual three-dimensional molecular cloud through absorption imaging. 
		The third dimension of the cloud along the imaging direction is integrated over,  which influences the \AW{density cross-}correlations measured.
		Furthermore, the speckle grains are inhomogeneous. While they have their long axis along the direction of imaging,  the extension of the molecular cloud is still larger than the length of the speckle grains as about two grains fit into the cloud along the imaging direction. 
		Moreover, the positioning of the grains along the imaging direction is random. Thus, some lie only inside a small part of the cloud, while others may be entirely immersed in the cloud. Due to the two-dimensional image, it is unclear how many speckle grains are causing the measured density fluctuations. 
		Some speckle grains are thus not perceived, systematically changing the correlation properties measured. 
		In other cases, independent speckle grains may lie very close to each other and be perceived as a single grain in density-density cross-correlated images. 
		This leads to a systematic error in which only a part of the speckle is perceived. 
		Thus, the effects of the optical speckle on the molecular density distribution in the absorption image are reduced, causing an apparently larger decorrelation angle. 
		
\section{Conclusions}
		In conclusion, we have constructed a setup to generate time-dependent disorder with constant correlation length but \AW{time-controlled correlation angle}.
		This method allows adjusting all relevant parameters of an optical disorder potential, including the potential depth  and the time scales of the disorder dynamics, as required. 
		We characterized a BEC in varying (but static) disorder realizations to extract the disorder correlation angle. 
		Our results confirm that the density-density cross-correlations obtained from different quantum-gas images used as a probe for the disorder potential show the same behavior as the intensity distributions we recorded \textit{ex-situ}.
		However, the absorption-imaging properties of the quantum gas limit the \textit{in-situ} method, leading to apparently larger correlation angles.
		
\AW{ An appealing extension of our scheme is a continuous evolution of the disorder, where the disorder realization changes while one and the same quantum gas is trapped. 
For a diffuser continuously rotating at angular frequency $\omega_\mathrm{r}$, we can define the correlation time as the temporal counterpart of the correlation length via
	\begin{equation} \label{eq:correlation_time}
		\tau = \frac{\alpha_{corr}}{\omega_\mathrm{r}}.
	\end{equation}
	The correlation time determines when a strong decorrelation is achieved, and the disorder potential loses resemblance to the original intensity pattern. 
		From  \Cref{eq:correlation_time}, using typical experimental parameters, we can also quantify the correlation times accessible in our setup.
		For angular velocities of up to $\ang{2100}\,\SI{}{\second^{-1}}$,  we find correlation times $\tau > \SI{543}{\micro \second}$. This time scale is bigger than the corresponding minimum time scale of the quantum gas given by the inverse chemical potential $h/\mu = \SI{190}{\micro \second}$. Hence, the gas can always adapt to the speckle potential’s changes for this largest chemical potential. 
		For a smaller chemical potential, however, we can realize an experimental situation where the many-body state of the gas cannot follow the fast changes of the speckle potential.
		Because of the systematic errors in the measurement of correlation angles via density cross-correlations of the molecular absorption pictures, we assume the real correlation angle to be smaller than \ang{1.14}, which leads to shorter correlation times. However, we can not resolve this any finer with the current technical constraints. With the \textit{ex-situ} determined correlation angle of \ang{0.49}, the minimal correlation time would be roughly \SI{233}{\micro \second}.		}
		This method of continuously transforming one disorder realization into another with an adjustable correlation time resembles naturally fluctuating disordered systems. 
		It is, therefore, well suited for future simulation of quantum dynamics, including, for example, the motion of ions in a fluctuating polymer background \cite{Aziz.2018},  as found in Lithium-polymer batteries. 
		
		Complementary methods to model time-dependent disorder, e.g.,  ionic transport through biomembranes \cite{Ratner.1989, Druger.1988}, use dynamic-disorder hopping models \cite{Druger.1990, Wang.2019, Wang.2020}, where the disorder instantly changes after certain renewal times.  While both models are expected to agree for short renewals and long observation times,  this is not the case for the short-time behavior, i.e., when individual renewals and their time scales matter.
		These situations could be realized using digital mirror devices \cite{Choi.2016} or spatial light modulators \cite{White.2020}, allowing for experiments with instantly changing potentials and different random patterns.
		
		Our characterization measurements point towards realizing transport studies in time-dependent and controllable disorder in strongly interacting quantum gases, extending the initial work on superfluid transport in disorder \cite{Chen2008}, or in interacting thermal clouds.
		An interesting question concerns the role of localized, noninteracting quantum systems in disordered potentials \cite{Bouyer.2010, Moratti.2012, Semeghini.2015} when the disorder is rendered time-dependent.  While in the asymptotic limit, localization is expected to cease in time-dependent disorder
		\cite{Ponte.2015, Abanin.2016, Lazarides.2015},  future experiments in our system might reveal how dynamics ensue in an originally localized quantum system, how the time scales and mobile fraction depend on the properties of the disorder potential, and if the microscopic time scales responsible for breaking the localization can be identified.
		
	\section*{Data availability statement}
		The data of this work is openly available at the following URL/DOI: \url{https://zenodo.org/records/10201108}.
				
	\section*{Acknowledgements}
		We acknowledge helpful discussions with A. Pelster and G. Orso, and discussions in the context of DAAD-CAPES.
		This work was supported by the Deutsche Forschungsgemeinschaft (DFG, German Research Foundation) via	the Collaborative Research Center SFB/TR185 (Project
		No. 277625399). We acknowledge the support of the Coordena\c{c}\~{a}o de Aperfei\c{c}oamento de Pessoal de N\'{i}vel Superior (\textsc{capes}) and the Deutscher Akademischer Austauschdienst (\textsc{daad}) under the bi-national joint program \textsc{capes-daad probral} Grant number 88887.627948/2021-00.
		J.K. was supported by the Max Planck Graduate Center with the Johannes Gutenberg-Universität Mainz (MPGC).

	\bibliography{paper_dynamic_speckle_characterization}{}

	\clearpage
	\section*{Appendix}
	\subsection*{Relation between optical and density correlations}
	\label{sec:dens_corr}
	In particular, while the behavior of the correlation angle of the optical speckle pattern can be deduced from direct measurements and knowledge of the rotation angle of the diffusive plates, it is not obvious that the same correlation properties can be determined from measurements of the molecular density distribution in the trap when superposing the speckle disorder. 
	Therefore, we relate the speckle-induced density corrugations detected in \textit{in-situ} absorption images with the independently measured correlation properties of the original optical speckle potential.
	For the analysis, it is of decisive importance to know the composition of the recorded absorption images. 
	These images comprise the signal of the disorder-trapped molecules and the characteristics of the imaging system.
	The optical system's influence is assumed to be known and can be independently measured. 
	These two contributions can be combined using the isoplanatic imaging equation yielding the final camera image as \cite{Zhuang.2016, Born.2019}
	\begin{equation}
		I=PSF\conv S  \label{eq:imaging equation}
	\end{equation}
	with the convolution operator $\conv$.
	It states that an image ($I$) is the convolution of the setup's point-spread function ($PSF$) with the signal ($S$) of an imaged object. 
	In our case, the images are given by the absorption images of the disorder-perturbed molecular clouds recorded, the $PSF$ results from the imaging system, which is limited by our camera resolution and therefore has the shape of single pixels,  and the signal ($S$) is the fluorescence from the speckle-disordered molecular cloud itself, see \cite{Katz2014, Meitav.2016}.
	
	The cross correlation \mbox{(denoted as $\cc$)} of two images $I_1$ and $I_2$ for  different rotation angles can be rewritten via the Fourier transform operator $\FT$ as
	\begin{equation*}
		\FT(I_1\cc I_2)=\overline{\FT(I_1)}\cdot \FT(I_2).\\
	\end{equation*}
	Using $I=I_1 \cc I_2 \text{ into } S=S_1 \cc S_2$ and \Cref{eq:imaging equation} yields \cite{Katz2014, Bertolotti2012, Dainty.1975}
	\begin{eqnarray}
		\FT(I_1\cc I_2)(x^\prime)	&= \overline{\FT(PSF\conv S_1)}(x^\prime)\cdot \FT(PSF\conv S_2)(x^\prime) \nonumber \\
	\end{eqnarray}	
	with control variable $x^\prime$. 
	After transforming back and employing the convolution theorem, this results in 
	\begin{equation}
		(I_1 \cc I_2)(x^\prime) = (PSF \cc PSF)(x^\prime) \conv (S_1\cc S_2)(x^\prime). \label{eq:ImageCorrelation}
	\end{equation}
	
	\noindent
	In the following, we will be interested in the maximum intensity-intensity correlation $I_{\Theta}(x_0)$ for quantifying the cross correlation of the two images $I_1$ and $I_2$ resulting from speckle realizations with relative rotation angle $\Theta$. 
	That allows us to reduce \Cref{eq:ImageCorrelation} to a simpler expression for the maximum intensity-intensity correlation at a given rotation angle $\Theta$ as
	\begin{eqnarray}
		I_{\Theta}(x_0)&=\int^{\infty}_{\infty} \mathrm{AC}(PSF(x^\prime)) \cdot S_{\Theta}(x_0-x^\prime) dx^\prime, \label{eq:AutoCorrelationFunction}
	\end{eqnarray}
	where $x_0$ is the initial position and $\mathrm{AC}(\cdot)$ denotes the auto-correlation function.
	The $PSF$ is the same for all images because the basic optical configuration remains the same for all images and rotations. Therefore, its autocorrelation function also contributes a constant factor to the integral, and \Cref{eq:AutoCorrelationFunction} leads to 
	\begin{equation}
		I_{\Theta}(0)\propto S_{\Theta}	.	
	\end{equation}
	\noindent
	This means that the image correlations reflect the speckle pattern correlations except for a constant factor originating from the imaging system. With this, we can extract the correlation properties of the speckles from the \textit{in-situ} density distributions of the molecular clouds.
	\\
	The correlation angle is calculated as the angle where the intensity-intensity correlation function $I_\Theta(x_0)$ has decayed to half the maximum value.
	One crucial point is that this only allows deducing and comparing the amount of correlations between two images.
	The opposite, i.e., inferring the complete information of the whole image from the correlation function, is not possible, especially since information on the molecular clouds can not be reconstructed. 
	Reconstructions of the original images are possible but computationally intensive and usually require more detailed knowledge of the specific PSF \cite{Shi:19, JulianC.Christou.1995, Shi.2020, He.2019}. 
	To reconstruct the original speckle pattern (or density distribution) using this setup, one would need the exact point-spread function and not just a rough estimate. 
	Nevertheless, our technique allows comparing the clouds for the molecular density. 
	The inversion of the density distribution's minima to the maxima of the speckle does not change the differences in the correlation between two pictures evaluated the same way.  \\

\end{document}